\RequirePackage{fix-cm}

\documentclass[twocolumn]{svjour3}

\usepackage{url}
\usepackage{color}
\usepackage{paralist}
\usepackage{fancyvrb}
\usepackage{epsfig}
\usepackage{subfig}
\usepackage{amsmath}
\usepackage[nomove]{cite}
\usepackage{tabularx}
\usepackage[table]{xcolor}
\usepackage{epstopdf}
\usepackage[numbers]{natbib}
\usepackage{booktabs} 
\usepackage{balance}

\newcommand{\q}[1]{\lq\lq{}{}#1\rq\rq{}{}}

\newcommand{\argmax}{\operatornamewithlimits{argmax}}
\newcommand{\argmin}{\operatornamewithlimits{argmin}}

\begin{document}\sloppy

\title{
Tracking the History and Evolution of Entities: Entity-centric Temporal Analysis of Large Social Media Archives
}
\titlerunning{Tracking the History and Evolution of Entities: Entity-centric Temporal Analysis of Large Social Media Archives}

\author{
    Pavlos Fafalios$^\star$ \and
    Vasileios Iosifidis$^\star$ \and
    Kostas Stefanidis$^\dagger$ \and
    Eirini Ntoutsi$^\star$ }

\authorrunning{Fafalios et al.}

\institute{
    $^\star$L3S Research Center, University of Hannover, Germany\\
    \email{\{fafalios, iosifidis, ntoutsi\}@l3s.de} \and \\
    $^\dagger$Faculty of Natural Sciences, University of Tampere, Finland\\
    \email{kostas.stefanidis@uta.fi} }

\date{Received: date / Accepted: date}

\maketitle

\begin{abstract}

How did the popularity of the Greek Prime Minister evolve in 2015? How did the predominant sentiment about him vary during that period? Were there any controversial sub-periods? What other entities were related to him during these periods? 

To answer these questions, one needs to analyze archived documents and data about the query entities, such as old news articles or social media archives. In particular, user-generated content posted in social networks, like Twitter and Facebook, can be seen as a comprehensive documentation of our society, and thus meaningful analysis methods over such archived data are of immense value for sociologists, historians and other interested parties who want to study the history and evolution of entities and events.
To this end, in this paper we propose an {\em entity-centric} approach to analyze social media archives and we define measures that allow studying how entities were reflected in social media in different time periods and under different aspects, like popularity, attitude, controversiality, and connectedness with other entities. A case study using a large Twitter archive of four years illustrates the insights that can be gained by such an entity-centric and multi-aspect analysis.

\keywords{
    Social Media Archives \and
    Entity Analytics \and
    Entity Linking \and
    Sentiment Analysis
}

\end{abstract}

\section{Introduction}
Social networking services have now emerged as central media to discuss and comment on breaking news and noteworthy events that are happening around the world. In Twitter, for example, every second around 6,000 tweets are posted, which corresponds to over 350,000 tweets per minute, 500 million tweets per day and around 200 billion tweets per year\footnote{\url{http://www.internetlivestats.com/twitter-statistics/} (August 30, 2018)}.

Such large amount of user-generated content produced continuously in social media is considered of immense historical value for future generations \cite{bruns2016twitter}. However, although there are initiatives that aim to collect and preserve social media archives, like the Twitter Archive at the Library of Congress \cite{zimmer2015twitter}, the absence of meaningful access and analysis methods still remains a major hurdle in the way of turning such archives into useful sources of information for historians, journalists and other interested parties \cite{bruns2016twitter}. 

When exploring archived data, analysts are not interested in the documents per se, but instead they want to see, compare, and understand the behavior of (and trends about) entities, like companies, products, politicians, athletes, celebrities, or music bands, thus calling for \textit{entity-level} analytics over the archived data \cite{weikum2011longitudinal}.

In this paper, we propose an {\em entity-centric} approach to analyze social media archives. Our approach allows tracking of how entities are reflected in a collection of user-generated content (e.g., tweets) in different time periods, and how such information evolves over time and also with respect to other entities. 
Specifically, we propose a multi-aspect description of an entity in terms of its
{\em popularity} (how much discussion it generates), {\em attitude}
(predominant sentiment towards the entity), {\em sentimentality} (magnitude of
sentiment towards the entity), {\em controversiality} (whether there is a consensus about the sentiment towards the entity), {\em connectedness} to another
entity (how strong is its connection to another entity), and {\em network} (strongly connected entities). We propose measures that capture all these aspects in a given time period (e.g., day, week, or month). 
A distinctive characteristic of our approach is that it does not rely on service-specific labels (like \#hashtags and @mentions), rather it exploits \textit{entity linking} \cite{shen2015entity} and thus can be applied over any type of time-annotated texts.

We examine the insights gained by the proposed measures on a large collection of billions of tweets spanning a period of 4 years (Jan 2013 - Jan 2017). Such analytics enable to answer questions like the following: 

\begin{itemize}
\item   {\em How did the popularity of Greek Prime Minister, Alexis Tsipras, evolve in 2015? Were there any \q{outlier} periods, i.e., periods of extremely high or low popularity? What were the entities discussed in social media together with Alexis Tsipras during these periods?
How did the \q{connectedness} of Alexis Tsipras with Vladimir Putin evolve in 2015?
}

\item   {\em How did the predominant sentiment about Donald Trump and Hillary Clinton vary during 2016? Were there any \q{controversial} time periods related to these two politicians, i.e., time periods in which there were many positive and negative tweets? 
What other entities were discussed together with Donald Trump and Hillary Clinton in tweets with predominant positive or negative sentiment?}
\end{itemize}

In a nutshell, this paper makes the following contributions:

\begin{itemize}
\item We propose a set of measures for capturing important entity features in a given time period. A sequence of such captures comprises a multi-variate time series in which each point is a multi-aspect description of the entity at a certain time period. We demonstrate the usefulness of our approach through illustrative examples.
      
\item We provide an open source Apache Spark library for computing the proposed measures efficiently.

\item We analyze a large Twitter archive spanning 4 years and containing billions of tweets and make publicly available the entity and sentiment annotations of this archive. This dataset can foster further research in related topics like topic evolution, entity recommendation, and concept drift.
\end{itemize}

This paper is an extension of \cite{fafalios2017tpdl}. The major changes include: i) an extensive survey of the related literature, ii) a new family of time-related measures (\textit{Entity-Time Measures}), iii) an extension of the entity-relation measures with new measures for identifying the networks with positive or negative sentiment of a given entity (\textit{Positive and Negative k-Networks}), iv) an extension of the case study with results related to the new measures.

The rest of this paper is organized as follows: 
Section \ref{sec:rw} provides the required background and related works. Section \ref{sec:approach} motivates and introduces the proposed measures. Section \ref{sec:library} describes a library for the distributed computation of the measures. Section \ref{sec:casestudy} presents the results of a case study. Finally, Section \ref{sec:conclusion} concludes the paper and identifies interesting directions for future research.


\section{Background and Related Literature}
\label{sec:rw}

\subsection{Entity and Sentiment Annotations}
\label{subsec:bg}

Our analysis is based on two different types of annotations applied in short texts from social media archives: {\em entity linking} and {\em sentiment analysis}. 

\subsubsection{Entities and Entity Linking}
Following Chen's definition \cite{chen1976entity}, an entity is \textit{\q{a thing which can be distinctly identified}}. In our problem, an entity has a Web identity expressed through a unique URI \cite{heath2011linked}.  
This does not only include persons, locations, organizations, etc., but also events (e.g., {\em US 2016 presidential election}) and general concepts (e.g., {\em democracy} or {\em abortion}).
A knowledge base contains information about a set of entities, like properties or relations with other entities. This information is described using one or more ontologies/vocabularies \cite{chandrasekaran1999ontologies}. DBpedia, for instance, is a cross-domain knowledge base derived from Wikipedia that makes use of the DBpedia Ontology for describing information about its entities \cite{lehmann2015dbpedia}.

Entity linking is the task of automatically identifying entity mentions in a piece of text and resolving them to their corresponding entries in a reference knowledge base  \cite{shen2015entity}. 
For example, given the text {\em\q{Obama visited Cuba}} and the reference knowledge base DBpedia, an effective entity linking system should link the text \q{Obama} to the former USA president Barack Obama (\url{http://dbpedia.org/resource/Barack_Obama}), and the text \q{Cuba} to the country Cuba (\url{http://dbpedia.org/resource/Cuba}). 
For each annotation, an entity linking system also provides a {\em confidence score} representing the confidence that the corresponding mention has been correctly disambiguated. 
The survey by Shen et al. \cite{shen2015entity} presents a thorough overview and analysis of the main approaches to entity linking, and discuss various applications as well as the evaluation of entity linking systems.
In our case studies, we used the system Yahoo FEL \cite{blanco2015fast} which has been specially designed for linking entities from short texts to DBpedia/Wikipedia. 

\subsubsection{Sentiment Analysis}
Sentiment analysis refers to the problem of assigning a sentiment label (e.g., positive, negative) or sentiment score to a document \cite{pang2008opinion}.
We opt for the latest and we use SentiStrength, a robust tool for
sentiment strength detection on social web data \cite{thelwall2012sentiment}.
SentiStrength assigns both a positive and a negative score (since both types of sentiment can occur simultaneously in a text). The strength score of a positive sentiment ranges from +1 (not positive) to +5 (extremely positive). Similarly, negative sentiment strength scores range from -1 (not negative) to -5 (extremely negative).
For example, given the text {\em\q{I love you but hate the current political climate}}, SentiStrength provides the positive sentiment score +3 and the negative sentiment score -4. 

\vspace{2mm} 
In Section \ref{subsec:dataset} we report evaluation results regarding the accuracy of Yahoo FEL and SentiStrength.

\subsection{Related Works}
\label{subsec:rw}
The availability of web-based application programming interfaces (APIs) provided by popular social media services like Twitter and Facebook, has led to an \q{explosion} of techniques, tools and platforms for social media analytics.
Batrinca and Treleaven \cite{batrinca2015social} surveys analytics tools for social media
as well as tools for scraping, data cleaning and sentiment analysis on social media data.
There is also a plethora of works on exploiting social media for a variety of tasks, like
opinion summarization \cite{meng2012entity},
event and rumor detection \cite{atefeh2015survey,qazvinian2011rumor},
topic popularity and summarization \cite{ardon2011spatio,yao2016tweet},
information diffusion \cite{guille2013information},
popularity prediction \cite{saleiro2016learning},
and reputation monitoring \cite{amigo2014overview}.
Furthermore, social media is exploited by research communities for research and experimentation in a variety of research problems. Examples include the {\em Making Sense of Microposts} series of workshops \cite{rizzo2015making,rizzo2016making}, or the {\em Sentiment Analysis in Twitter} tasks of the International Workshop on Semantic Evaluation \cite{nakov2016semeval,rosenthal2017semeval}.
Below, we describe works related to the {\em temporal} analysis of topics and entities in social media.

Stefanidis and Koloniari \cite{DBLP:conf/cikm/StefanidisK14} propose a query-answering
framework to allow entity search in social networks by exploiting the underlying social graph and temporal information. To satisfy the varying search needs, the framework includes a time-aware query model and a corresponding logical algebra. To deal with the temporal aspect, the authors adopt an annotated graph model that incorporates time by associating each element in the graph with its temporal information. The nodes (representing users and objects) and the edges between them (representing social relationships) have a label that indicates their valid time. The proposed query model allows for time-dependent queries that exploit time explicitly by using it as a hard constraint to filter out irrelevant results.

Ardon et al. \cite{ardon2011spatio} perform a spatiotemporal analysis of tweets, investigating the time-evolving properties of the subgraphs formed by the users discussing each topic. The focus is on the network topology formed by follower-following links on Twitter and the geospatial location of the users. The authors investigated the effect of initiators on the popularity of topics and find that users with a high number of followers have a strong impact on popularity. They also showed that topics become popular when disjoint clusters of users discussing them begin to merge and form one giant component that grows to cover a significant fraction of the network. 

Bruns and Stieglitz \cite{bruns2013towards} introduce a catalogue of metrics for analyzing hashtag-based communication on Twitter, with particular focus on hashtagged Twitter conversations. The proposed metrics can be categorized into: metrics that examine the total activity and visibility of individual participants, metrics that establish the temporal flow of conversation and of specific forms of conversation, and metrics that combine the activity of the users and the flow of conversations to examine the relative contributions of specific user groups in different time points.

Saleiro and Soares \cite{saleiro2016learning} tackle the problem of predicting entity popularity
on Twitter based on the news cycle. The authors apply a supervised learning
approach and extract four types of features (signal, textual, sentiment, and semantic) which are used to predict whether the popularity of a given entity will be high or low in the following hours. The results of an experimental evaluation showed that news perform better on predicting entity popularity on Twitter when they are the primary information source of the event, in opposition to events such as live TV broadcasts, political debates or football matches.

Celik et al. \cite{celik2011learning} investigate whether semantic relationships
between entities can be learned by analyzing microblog posts published on Twitter. The authors developed a relation discovery framework that allows for the detection of typed relations that may have temporal dynamics. 
The evaluation results showed that co-occurrence based strategies allow for high precision and perform particularly well for relations between persons and events. Our entity-to-entity connectedness scores are also based on entity co-occurrences
(more in Section \ref{sec:approach}). The authors also analyzed the performance in
learning relationships that are valid only for a certain time period and
revealed that Twitter is a suitable source for this type of relationships because
it allows the discovery of trending topics with high accuracy and low delay. 

Ren et al. \cite{ren2013personalized}  consider the task of time-aware tweets summarization exploiting user's history and collaborative social influences from social circles. The authors propose a time-aware user behavior model, called Tweet Propagation Model, in which dynamic probabilistic distributions over interests and topics are inferred. 
In the same context, Zhao et al. \cite{zhao2013timeline} study how to incorporate social attention in the generation of timeline summaries. Given a topic, the authors propose learning users' collective interests in the form of word distributions from Twitter which are subsequently incorporated into a unified framework for timeline summary generation.
In a similar problem, Chang et al. \cite{chang2013towards} introduce the task of Twitter context summarization, which generates a succinct summary from a large but noisy Twitter context tree. The authors study how user influence models, which project user interaction information onto a Twitter context tree, can help Twitter context summarization within a supervised learning framework.

Regarding more recent works on timeline summarisation, Yao et al. \cite{yao2016tweet} focus on how to select a small set of representative tweets to generate a meaningful timeline, which provides enough coverage for a given topical query. The proposed approach jointly models individual topical relevance and overall diversity within a probabilistic model. 
Chang et al. \cite{chang2016timeline} propose a framework called {\em Timeline-Sumy}, which consists of two main components: {\em episode detecting}, and {\em summary ranking}. Episode detecting aims to identify key episodes in a timeline, while  summary ranking ranks the social media posts in each episode via a learning-to-rank approach. 

Finally, Li and Cardie \cite{li2014timeline} propose an unsupervised framework for creating a chronological list of a user's personal important events. The authors introduce a non-parametric multi-level Dirichlet Process model to recognize four types of tweets: personal time-specific, personal time-general, public time-specific, and public time-general. These tweets, in turn, are used for further personal event extraction and timeline generation.

To our knowledge, our work is the first that models multi-aspect {\em entity-centric} analytics for social media archives, by combining automatically extracted entities with sentiment information expressed in the tweets. The proposed measures capture the multi-aspect behavior of an entity in different time periods and can be exploited in a variety of tasks, like entity evolution, event detection, and entity recommendation. In addition, our approach does not rely on service-specific labels (likes hashtags) and thus can be applied over any type of time-annotated short texts.


\section{Multi-aspect Entity Measures}
\label{sec:approach}

\subsection{Motivation}
\label{subsec:motivation}

According to Weikum et al. \cite{weikum2011longitudinal}, when exploring archived data, like old web archives, analysts prefer to deal with semantically rich entities like people, places, organizations, and ideally relationships among them, instead of documents containing such references. The authors envision a system that should support a wide spectrum of analytical tasks that span the text, entity and time dimensions, such as identification of salient entities for different subsets of an archive, entity-to-entity co-occurrences, or detection of interesting time points or periods for a given entity. 
In addition, to preserve Twitter as a historical source, Bruns and Weller \cite{bruns2016twitter} suggest that important events should be monitored while systems should offer the possibility to collect tweets for single events in order to document important background information or other contextual information (like related entities). 

Considering the above, we propose a set of {\em entity-centric} measures that allow studying how entities (including events) are reflected in social media in different time periods and under different aspects. We propose a multi-aspect description of an entity in terms of the following aspects (computed for a given time period, like a specific day, week, or month):

\begin{itemize}
    \item {\em entity popularity} (how much discussion it generates)
    \item {\em entity attitude} (predominant sentiment towards the entity)
    \item {\em entity sentimentality} (magnitude of sentiment towards the entity)
    \item {\em entity controversiality} (whether there is a consensus about the sentiment towards the entity)
    \item {\em entity-to-entity connectedness} (how strong is its connection to another entity)
    \item {\em entity network} (strongly connected entities)
\end{itemize}

These time-dependent entity features can facilitate research in a plethora of related  problems, including prediction tasks \cite{yu2012survey,saleiro2016learning} (by exploiting \textit{popularity}, \textit{attitude} and \textit{sentimentality}), controversy detection \cite{garimella2018quantifying} (by exploiting \textit{controversiality}), time-aware entity relatedness \cite{mohapatra2018timeaware} (by exploiting {\em entity-to-entity connectedness}), and time-aware entity recommendation \cite{zhang2016probabilistic} (by exploiting {\em entity network}).

Below, we formally introduce the proposed measures by classifying them into three categories: i) {\em single-entity measures}, ii) {\em entity-time measures}, and iii) {\em entity-relation measures}.

\subsection{Single-Entity Measures}
\label{subsec:singleMeasures}

First, let $C$ be a collection of short texts (e.g., tweets)
covering the time period $T = [t_s, t_e]$ (where $t_{s}, t_{e}$ are
two different time points with $t_{s} < t_{e}$), and let $U$ be the
total set of users who posted these texts. Let also $E$ denote a
finite set of entities, e.g., all Wikipedia entities.

\subsubsection*{Popularity}
Let $e \in E$ be a given entity and $T_i \subseteq T$ a given time
period. Let also $C_i \subseteq C$ be the collection of short texts
posted during $T_i$. The popularity of $e$ during $T_i$ equals to
the percentage of {\em texts} mentioning $e$ during that period.
Formally:
\begin{equation}
\label{eq:pop1} 
\centering
popularity_c(e, T_i) = \frac{|C_{e, i}|}{|C_i|}
\end{equation}
where $C_{e, i} \subseteq C_i$ denotes the set of texts mentioning
$e$ during $T_i$.

Using the above measure, an entity can be very popular
even if it is discussed by a few users but in a large
number of texts. A more fine-grained indication of popularity is
given by the number of different users discussing the entity.
In that case, if $u_c \in U$ denotes the user who posted the text
$c$, the popularity of an entity $e \in E$ during $T_i$ can be
defined as the percentage of different {\em users} discussing $e$ during that period, i.e.:
\begin{equation}
\label{eq:pop2} 
\centering
popularity_u(e, T_i) = \frac{|\cup_{c \in C_{e,i}}{u_c}|}{|\cup_{c \in C_{i}}{u_c}|}
\end{equation}

\noindent 
We can now combine both aspects (percentage of {\em texts} and {\em users}) in one popularity score:
\begin{equation}
\label{eq:popAll} 
\begin{split}
popularity_{c,u}(e, T_i) = & \\  ~~~~~popularity_c(e, T_i)
\cdot popularity_u(e, T_i)
\end{split}
\end{equation}
An entity has now a high popularity score if it is discussed in
many tweets and by many different users.

\subsubsection*{Attitude and Sentimentality}
We use two measures (proposed by Kucuktunc et al. \cite{kucuktunc2012large} for the case of questions and answers)
for capturing a text's {\em attitude} (predominant sentiment) and {\em
sentimentality} (magnitude of sentiment). First, for a text $c \in
C$, let $s^+_c \in [1,5]$ be the text's positive sentiment score and
$s^-_c \in [-5,-1]$ be the text's negative sentiment score
(according to SentiStrength, c.f. Section \ref{subsec:bg}).
The attitude of a text $c$ is given by $\phi_c = s^+_c + s^-_c$ (i.e., $\phi_c \in [-4,
4]$) and its sentimentality by $\psi_c = s^+_c - s^-_c - 2$ (i.e., $\psi_c \in [0, 8]$).

We now define the {\em attitude} of an entity $e$ in a time period $T_i$ as the average
attitude of texts mentioning $e$ during $T_i$. Formally:
\begin{equation}
\label{eq:attitude} attitude(e, T_i) = \frac{\sum_{c \in C_{e,
i}}{\phi_{c}}}{|C_{e, i}|}
\end{equation}

Likewise, the {\em sentimentality} of an entity $e$ in a time period
$T_i$ is defined as the average sentimentality of texts mentioning
$e$ during $T_i$:

\begin{equation}
\label{eq:sentimentality} sentimentality(e, T_i) = \frac{\sum_{c \in
C_{e, i}}{\psi_{c}}}{|C_{e, i}|}
\end{equation}

\subsubsection*{Controversiality}
An entity $e$ can be considered controversial in a time period $T_i$
if it is mentioned in plenty of both positive and negative texts.
First, let $C^+_{e, i}$ be the set of
texts mentioning $e$ during $T_i$ with strong positive attitude,
i.e., $C^+_{e, i} = \{c \in C_{e, i} ~|~ \phi_c \geq \delta \}$,
where $\delta \in [0, 4]$ is a strong attitude threshold (e.g.,
$\delta = 2.0$). Likewise, let $C^-_{e, i}$ be those with strong
negative attitude, i.e., $C^-_{e, i} = \{c \in C_{e, i} ~|~ \phi_c \leq -\delta \}$.
We now consider the following formula for entity
{\em controversiality}:
\begin{equation}
\label{eq:controv} 
\begin{split}
controversiality(e, T_i) = & \\ ~~~~~ \frac{|C^+_{e, i}| +
|C^-_{e, i}|}{|C_{e, i}|} \cdot \frac{min(|C^+_{e, i}|,~|C^-_{e, i}|)}{max(|C^+_{e, i}|,~|C^-_{e, i}|)} 
\end{split}
\end{equation}
Intuitively, a value close to 1 means that the probability of the
entity being \q{controversial} is high since there is a big
percentage of texts with strong attitude (first part of the formula)
and also there are many texts with both strong positive attitude and strong negative attitude (second part of the
formula).

\subsection{Entity-Time Measures}
\label{subsec:timePeriod}

By exploiting the single-entity measures, we can now compute important time sub-periods of granularity $\Delta$ (e.g., day, week or month) for a given entity in a given time-period.
For instance, given the entity {\em Barack Obama}, the time period {\em 2015} and the granularity {\em month}, we can find the top-3 months of 2015 of high or low Obama's popularity. Then, for a specific {\em month}, we can find the top-5 {\em days} of high or low Obama's popularity.
We define the following measures: 
\begin{itemize}
    \item Top-K time sub-periods of high/low popularity 
    \item Top-K time sub-periods of high/low attitude
    \item Top-K time sub-periods of high/low controversiality 
\end{itemize}

\subsubsection*{Top-K Time Periods of High/Low Popularity}

Given an entity $e$, a time period $T_i$ and a granularity $\Delta$, the {\em top-$k$ time periods of high popularity} of $e$ during $T_i$ is the set of $k$ time (sub-)periods of granularity $\Delta$ with the highest entity popularity score (cf. Formula \ref{eq:popAll}). 
Let first $T_{i, \Delta}$ be the set of all time (sub-)periods of granularity $\Delta$ covering the time-period $T_i$ (for example, all days in a month).
Now, the top-K time (sub-)periods of high popularity of $e$ during $T_i$ can be defined as:
\begin{equation}
\label{eq:topKpopHigh}
\begin{split}
k\text{-}HighPopularPeriods(e, T_i, \Delta) = & \\ ~~~~~  \argmax_{\substack{T' \subseteq T_{i,\Delta}, ~|T'| = k}}{~\sum_{t \in T'}{popularity_{c,u}(e, t)}}
\end{split}
\end{equation}

Likewise, the set of top-K time (sub-)periods of low popularity is defined as:
\begin{equation}
\label{eq:topKpopLow}
\begin{split}
k\text{-}LowPopularPeriods(e, T_i, \Delta) = & \\ ~~~~~  \argmin_{\substack{T' \subseteq T_{i,\Delta}, ~|T'| = k}}{~\sum_{t \in T'}{popularity_{c,u}(e, t)}}
\end{split}
\end{equation}

\subsubsection*{Top-K Time Periods of High/Low Attitude}

By exploiting the attitude measure (cf. Formula \ref{eq:attitude}), we can find time periods of high or low entity attitude. 
Given an entity $e$, a time period $T_i$ and a granularity $\Delta$, the {\em top-$k$ time periods of high attitude} of $e$ during $T_i$ is the set of $k$ time (sub-)periods of granularity $\Delta$ with the highest entity attitude score. Formally:
\begin{equation}
\begin{split}
k\text{-}HighAttitudePeriods(e, T_i, \Delta) = & \\ ~~~~~  \argmax_{\substack{T' \subseteq T_{i,\Delta}, ~|T'| = k}}{~\sum_{t \in T'}{attitude(e, t)}}
\end{split}
\end{equation}

\vspace{2mm}
Likewise, the set of  {\em top-$k$ time periods of low attitude} is defined as: 
\begin{equation}
\begin{split}
k\text{-}LowAttitudePeriods(e, T_i, \Delta) = & \\ ~~~~~  \argmin_{\substack{T' \subseteq T_{i,\Delta}, ~|T'| = k}}{~\sum_{t \in T'}{attitude(e, t)}}
\end{split}
\end{equation}

\subsubsection*{Top-K Time Periods of High/Low Controversiality}

Given an entity $e$, a time period $T_i$ and a granularity $\Delta$, the {\em top-$k$ controversial time periods} of $e$ during $T_i$ is the set of $k$ time (sub-)periods of granularity $\Delta$ with the highest entity controversiality score (cf. Formula \ref{eq:controv}). Formally:
\begin{equation}
\label{eq:topKcontrHigh}
\begin{split}
k\text{-}HighControversialPeriods(e, T_i, \Delta) = & \\ ~~~~~  \argmax_{\substack{T' \subseteq T_{i,\Delta}, ~|T'| = k}}{~\sum_{t \in T'}{controversiality(e, t)}}
\end{split}
\end{equation}

\vspace{2mm}
Likewise, the set of  {\em top-$k$ time periods of low controversiality} is defined as:
\begin{equation}
\label{eq:topKcontrLow}
\begin{split}
k\text{-}LowControversialPeriods(e, T_i, \Delta) = & \\ ~~~~~  \argmin_{\substack{T' \subseteq T_{i,\Delta}, ~|T'| = k}}{~\sum_{t \in T'}{controversiality(e, t)}}
\end{split}
\end{equation}

\subsection{Entity-Relation Measures}
\label{subsec:relationMeasure}

Here we define measures that quantify the degree of association (or connectedness) of a query entity with other entities mentioned in the same collection.

\subsubsection*{Entity-to-Entity Connectedness}
We define a {\em direct-connectedness} score between an entity $e \in E$
and another entity $e' \in E$ in a time period $T_i$, as the number
of texts in which $e$ and $e'$ co-occur within $T_i$. Formally:
\begin{equation}
\label{eq:conT} direct\text{-}connectedness(e, e', T_i) = \frac{|C_{e, i} \cap
C_{e', i}|}{|C_{e, i}|}
\end{equation}
Notice that the relation is not symmetric. We consider that if an
entity $e_1$ is strongly connected with an entity $e_2$, this does
not mean that $e_2$ is also strongly connected with $e_1$. 
For example, consider that {\em Alexis Tsipras} is mentioned in only 100 texts during $T_i$, {\em Barack Obama} in 1M texts, while 90 texts mention both entities. We notice that
{\em Barack Obama} seems to be a very important entity for {\em Alexis Tsipras} during $T_i$, since it exists in 90/100 of texts mentioning {\em Alexis Tsipras}. 
On the contrary, {\em Alexis Tsipras} seems not to be important for {\em Barack Obama} , since it exists in only 90/1M of texts mentioning {\em Barack Obama}.

Two entities may not co-occur in texts, but they may share many
common co-occurred entities. For example, both {\em Barack Obama}
and {\em Donald Trump} may co-occur with entities like {\em White
House}, {\em US Election} and {\em Hillary Clinton}.
For an input entity $e \in E$ and another entity $e' \in E$,
we define an {\em indirect-connectedness} score which
considers the number of {\em common entities} with which $e$
and $e'$ co-occur in a time period $T_i$:

\begin{equation}
\label{eq:conE}
\vspace{1mm}
\begin{split}
indirect\text{-}connectedness(e, e', T_i) = & \\ ~~~~~ \frac{|(\cup_{c \in C_{e, i}}{E_c}) \cap (\cup_{c \in C_{e', i}}{E_c})|}{|(\cup_{c \in C_{e, i}}{E_c})|}
\end{split}
\vspace{2mm}
\end{equation}
where $E_c \subseteq E$ is the entities mentioned in text $c$.
Also in this case, the relation between the two entities is not symmetric.

\vspace{2mm}
\subsubsection*{Entity $k$-Network}
This measure targets at finding a list of entities strongly connected to the query entity in a given time period $T_i$. First, we define a connectedness score between an entity $e \in E$ and a set of entities $E' \subseteq E$ within $T_i$, as the average direct-connectedness score of the entities in $E'$. Formally:
\begin{equation}
\begin{split}
connectedness(e, E', T_i) = & \\ ~~~~~ \frac{\sum_{e' \in E'}{direct\text{-}connectedness(e, e', T_i)}}{|E'|}
\end{split}
\end{equation}

The $k$-Network of an entity $e$ during $T_i$ is the set of $k$ entities $E' \subseteq E$ with the highest average connectedness score. Namely:
\begin{equation}
\begin{split}
k\text{-}Network(e,T_i)= & \\ ~~~~~  \argmax_{\substack{E' \subseteq E, ~|E'| =
k}} connectedness(e, E', T_i)
\end{split}
\end{equation}

In simple terms, the $k$-Network of an entity $e$ consists of the $k$ entities with the highest {\em direct-connectedness} scores.

\vspace{2mm}
\subsubsection*{Positive \& Negative $k$-Networks}

Based on the attitude of the texts mentioning two entities, we can compute the corresponding positive and negative $k$-Networks.
First, for an entity $e \in E$ and a time period $T_i$, the set of positive entities $E_{e, i}^+ \subseteq E$ can be defined as the set of entities co-occurring with $e$ during $T_i$ in texts with strong average positive attitude, i.e.:

\begin{equation}
\begin{split}
E_{e, i}^+ = \{ e' \in E ~|~ \frac{\sum_{c \in C_{e, i} \cap C_{e', i}}{\phi_c}}{|C_{e, i} \cap C_{e', i}|} \geq \delta \}
\end{split}
\end{equation}
where $\delta \in [0, 4]$ is a strong attitude threshold (e.g., $\delta = 2.0$).
Likewise, the set of entities co-occurring with $e$ during $T_i$  in texts with strong average negative attitude can be defined as:
\begin{equation}
\begin{split}
E_{e, i}^- = \{ e' \in E ~|~ \frac{\sum_{c \in C_{e, i} \cap C_{e', i}}{\phi_c}}{|C_{e, i} \cap C_{e', i}|} \leq -\delta \}
\end{split}
\end{equation}

\vspace{2mm}
Now, the positive and negative $k$-Networks of an entity $e$ in a time period $T_i$
can be defined as:
\begin{equation}
\begin{split}
k\text{-}Network^+(e, T_i)= & \\ ~  \argmax_{\substack{E' \subseteq E_{e, i}^+, ~|E'| = k}} connectedness(e, E', T_i)
\end{split}
\end{equation}

\begin{equation}
\begin{split}
k\text{-}Network^-(e, T_i)= & \\ ~  \argmax_{\substack{E' \subseteq E_{e, i}^-, ~|E'| = k}} connectedness(e, E', T_i)
\end{split}
\end{equation}

\subsection{Discussion}

The above presented measures capture the multi-aspect behavior of a
given entity at a certain time period.
In the long run,
a multi-variate time series is formed where each point
represents the multi-aspect description of the entity at a certain period in time.

An important characteristic of our approach is that we can support both
entity-specific queries referring to a single entity and
cross-entity queries involving more than one entities (e.g., a
category of entities). This is achieved through the {\em entity
linking} process in which entities are extracted from the texts and
are linked to knowledge bases like Wikipedia/DBpedia. In that way, we can collect a variety of
properties for the entities extracted from our
archive. This enables us to aggregate information and capture the
behavior of sets of entities. For example, by accessing DBpedia, we
can collect a list of German politicians,
derive their popularity and then compare it with that of another
set of entities.

In addition, the proposed measures can be easily computed by submitting queries on related knowledge bases that contain metadata and annotation information about a collection of archived documents or social media posts \cite{fafalios2017jcdl,fafalios2018ijdlSemLay,fafalios2018tweetskb}. This enables the production of time series at query-execution time, thereby allowing the answer of complex information needs (through structured SPARQL queries) as well as \q{on the fly} data integration (by exploiting other knowledge bases like DBpedia).


\section{Library for Computing the Measures}
\label{sec:library}

For computing the measures, we provide an Apache Spark library.
Apache Spark\footnote{\url{http://spark.apache.org/}} is a
cluster-computing framework for large-scale data processing. The
library contains functions for computing the proposed measures for a
given entity and over a specific time period. It operates over an
annotated (with entities and sentiments) dataset split per
year-month (the dataset should be in a simple CSV format). The
library is available as open
source\footnote{\url{https://github.com/iosifidisvasileios/Large-Scale-Entity-Analysis}}.

The time for computing the measures highly depends on the dataset volume, the used computing infrastructure
as well as the available resources and the load of the cluster at the analysis time.
The Hadoop cluster used in our experiments for analyzing a large Twitter archive of
more than 1 billion tweets consisted of 25 computer nodes with a total of 268 CPU cores and 2,688 GB RAM
(more about the dataset in the next section).
Indicatively, the time for computing each of the measures was on average less than a minute (without using any index, apart from the monthly-wise split of the dataset).


\section{Case Study: Entity Analytics on a Twitter Archive}
\label{sec:casestudy}

In this section, we first describe the results of the analysis and annotation
of a large Twitter archive.
Then, we present examples of case studies illustrating the insights gained
from the proposed measures.

\subsection{Annotating a Large Twitter Archive}
\label{subsec:dataset}

\begin{figure*}[t]
    \centering
    \includegraphics[width=5.6in]{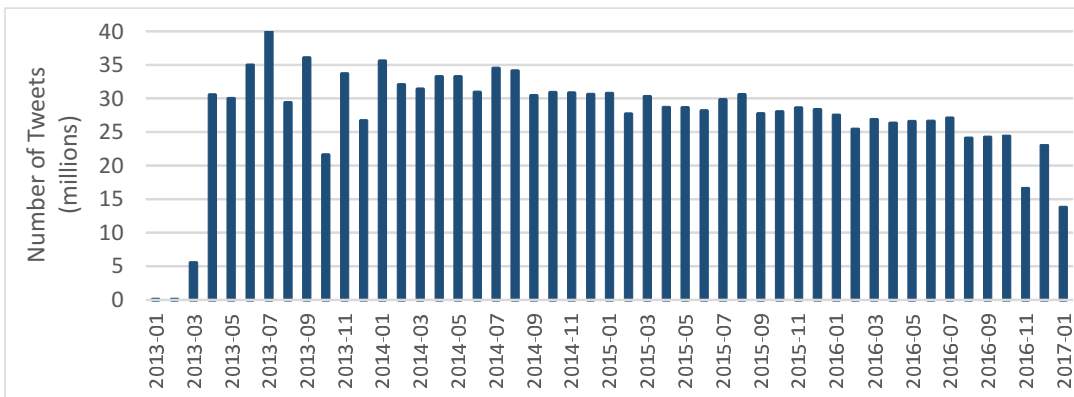}
    \vspace{-1mm}
    \caption{Number of tweets per month.}
    \label{fig:tweetsPerMonth}
\end{figure*}

We analyzed a large Twitter archive spanning 4 years (January 2014 - January 2017) and containing more than 6 billion tweets. The tweets were collected through the Twitter streaming API. Our analysis comprised the following steps: i) filtering (filtering out re-tweets, keeping only English tweets), ii) spam removal, iii) entity linking, and iv) sentiment analysis. The filtering step reduced the number of tweets to about 1.5 billion tweets (specifically, to 1,486,473,038 tweets). 
For removing the spam tweets, we trained a Multinomial Naive Bayes (MNB) classifier over the HSpam dataset \cite{sedhai2015hspam14}, which consists of tweets annotated as either spam or not. We applied the learned model to our dataset and removed all tweets classified as spam. This removed about 150 million tweets (around 10\% of the tweets). The final dataset consists of 1,335,324,321 tweets posted by 110,548,539 users. Figure \ref{fig:tweetsPerMonth} shows the number of tweets per month on the final dataset.

For the {\em entity linking} task, we used Yahoo FEL
\cite{blanco2015fast} with a confidence threshold score of -3.
In total, 1,390,286 distinct entities were extracted from the collection. For each extracted entity, its 
confidence score provided by FEL is also stored. Thereby, data consumers can select
suitable confidence ranges to consider, depending on the specific
requirements with respect to precision and recall. 
For {\em sentiment analysis}, we used SentiStrength
\cite{thelwall2012sentiment}. The average sentimentality of all
tweets is 0.92, the average attitude 0.2, while 622,230,607 tweets
have no sentiment (-1 negative sentiment and 1 positive sentiment).
Table \ref{tbl:attitude} shows the number of tweets per attitude
value.

\begin{table}
\centering
  \caption{Number of tweets per attitude value.}
  \label{tbl:attitude}
  \begin{tabular}{cr}
    \toprule
    Attitude value & Number of Tweets\\
    \midrule
    -4       & 2,234,887 (0.17\%)       \\
    -3       & 34,666,708 (2.60\%)      \\
    -2       & 68,812,370 (5.15\%)     \\
    -1       & 104,628,022 (7.84\%)      \\
    0        & 670,484,267 (50.2\%)     \\
    1        & 301,635,430 (22.6\%)    \\
    2        & 138,197,637 (10.3\%)     \\
    3        & 13,610,492 (1.02\%)     \\
    4        & 1,054,508 (0.08\%)      \\
  \bottomrule
\end{tabular}
\end{table}

\subsubsection*{Quality of entity annotations}

We used the ground truth dataset provided by the 2016 NEEL challenge of the 6th workshop on \q{Making Sense of Microposts} (\#Microposts2016)\footnote{\url{http://microposts2016.seas.upenn.edu/}} \cite{rizzo2016making} for evaluating the quality of the entity annotations produced by FEL. The dataset consists of 9,289 English tweets of 2011-2015. We considered all tweets from the provided training, dev and test files.
The results are the following: {\em Precision} = 86\%, {\em Recall} = 39\%, {\em F1} = 54\%.
We notice that FEL achieves high precision, however recall is low. The reason is that FEL did not manage to recognize several difficult cases, like entities within hashtags and nicknames, which are common in Twitter.
Nevertheless, FEL's performance is comparable to existing approaches \cite{rizzo2015making,rizzo2016making}.

\subsubsection*{Quality of sentiment annotations}

We evaluated the accuracy of SentiStrength using the ground truth datasets {\em SemEval2017} (Task 4, Subtask A)\footnote{\url{http://alt.qcri.org/semeval2017/task4/}} \cite{rosenthal2017semeval} and {\em TSentiment15}\footnote{\url{https://l3s.de/~iosifidis/TSentiment15/}} \cite{iosifidis2017large}. 
The {\em SemEval2017} dataset consists of 61,853 English tweets collected during the period 2013-2017 and labeled by human annotators as positive, negative, or neutral. We run the evaluation on all the provided training files (of 2013-2016) and on the 2017 test file.
SentiStrength achieved the following scores:  {\em  AvgRec = 0.54} (recall averaged across the positive, negative, and neutral classes \cite{sebastiani2015axiomatically}), {\em $F1^{PN}$ = 0.52} (F1 averaged across the positive and negative classes), {\em Accuracy} = 0.57. 
The performance of SentiStrength is good considering that this is a multi-class classification problem. The user can also achieve higher precision by selecting only tweets with high positive or negative SentiStrength score (e.g., $>$+2 for positive or $<-2$ for negative sentiment). 
Regarding the {\em TSentiment15} dataset, it consists of 2,527,753 English tweets collected during 2015, labeled as either positive or negative through semi-supervised learning \cite{iosifidis2017large}.
 SentiStrength achieved the following scores: {\em $F1^{PN}$} = 0.80, {\em Accuracy} = 0.91. 
Here we notice that SentiStrength achieves very good performance. 
To conclude, our evaluation on Twitter ground truth datasets showed that SentiStrength achieves good performance in sentiment annotation of tweets.

\subsubsection*{Dataset availability}
The annotated dataset is publicly available in CSV
format\footnote{\url{http://l3s.de/~iosifidis/tpdl2017/}}.
For each tweet the dataset includes the following information: ID, user
(encrypted), post date, extracted entities, positive and negative
sentiment values (the text of the tweets is not provided for
copyright purposes\footnote{\url{https://help.twitter.com/en/rules-and-policies/copyright-policy}}).

We make the dataset available so anyone interested can use it
together with the library (described in Section \ref{sec:library})
to extract the measures for any entity at the desired level of temporal granularity.
We believe that such efforts can foster further research in a variety of areas like {\em entity recommendation}, {\em entity summarisation} and {\em concept drift}.

\subsection{Case Studies}
\label{subsec:casestudies}

\subsubsection*{Entity Popularity}

\begin{figure}[t]
    \centering
    \includegraphics[width=3.0in]{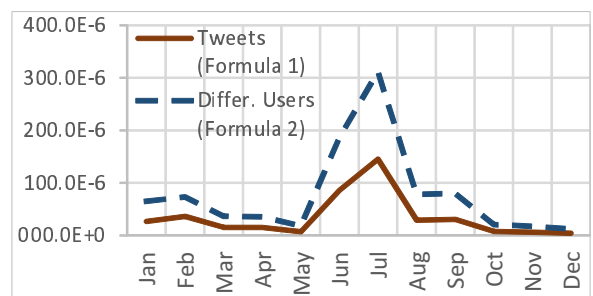}
    \caption{Popularity evolution of {\em Alexis Tsipras} in 2015.}
    \label{fig:popAlex}
\end{figure}

Figure \ref{fig:popAlex} shows the popularity of {\em Alexis Tsipras} (Greek prime minister) within 2015.
We notice that his popularity highly increased in July.
Indeed, in July 2015 the Greek bailout referendum was held following
the bank holiday and capital controls of June 2015.
This event highly increased the popularity of the Greek prime minister.
Moreover, by comparing the trend of the two different popularity scores (Formulas \ref{eq:pop1} and \ref{eq:pop2}),
we notice that, during June and July 2015,
the percentage of different users discussing about {\em Alexis Tsipras} increased
in bigger degree compared to the percentage of tweets,
implying that more people were engaged in the discussion.
As regards his {\em top-$K$ time periods of high/low popularity} (Formulas \ref{eq:topKpopHigh} and \ref{eq:topKpopLow}),
we notice that the top-3 months of high popularity in 2015 are [July, June, February], while the corresponding top-3 months of low popularity are [December, November, May].

\begin{figure}[t]
    \centering
    \includegraphics[width=2.6in]{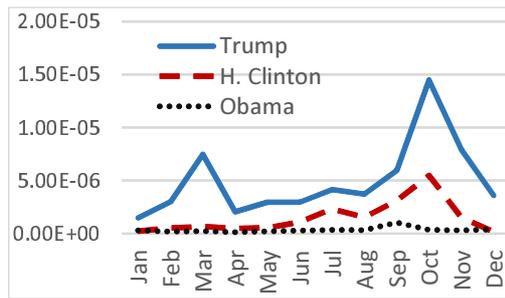}
    \caption{Popularity evolution of {\em Donald Trump}, {\em Hillary Clinton} and  {\em Barack Obama} in 2016.}
    \label{fig:popTrClOb}
\end{figure}

Likewise, we can compare the popularity of multiple entities within
the same time period. For example, Figure \ref{fig:popTrClOb} shows the popularity of {\em Donald Trump}, {\em Hillary
Clinton} and  {\em Barack Obama} within 2016 (according to Formula
\ref{eq:popAll}). We notice that {\em Donald Trump} is much more
popular in all months. We also notice that, in October 2016 the
popularity of {\em Donald Trump} and {\em Hillary Clinton} highly
increased compared to the other months. This is an indicator of
possible important events related to these two entities in October
2016 (indeed, two presidential general election debates took place
in that period).
The top-3 months of high popularity in 2016 are [October, November, March] for {\em Donald Trump}, [October, September, July] for {\em Hillary Clinton}, and [September, December, July] for {\em Barack Obama}.

\subsubsection*{Entity Attitude}

\begin{figure*}[t]
    \centering
    \includegraphics[width=6.85in]{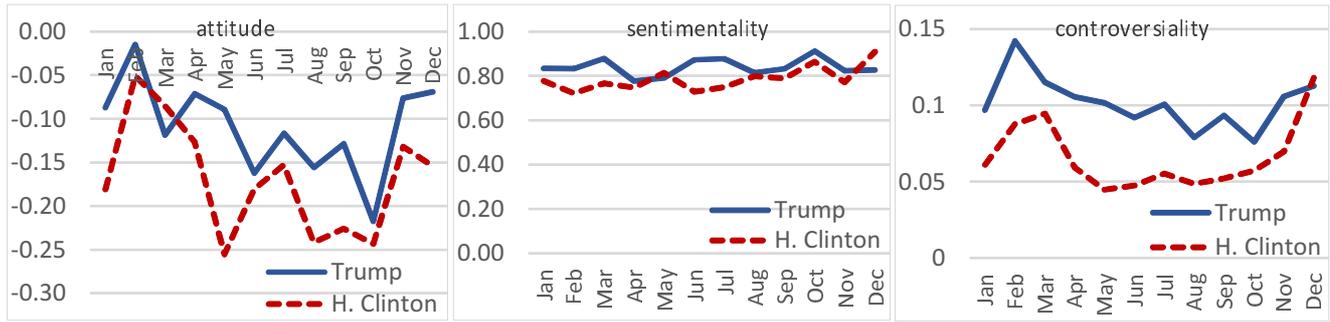}
    \caption{Evolution of attitude (left), sentimentality (middle), and controversiality (right) of {\em Donald Trump} and {\em Hillary Clinton} in 2016.}
    \label{fig:attSentContr}
\end{figure*}

Figure \ref{fig:attSentContr} (left) depicts the attitude
 of {\em Donald Trump} and {\em Hillary Clinton}
within 2016. We notice that both entities had constantly a negative
attitude, however that of {\em Hillary Clinton} was worse in almost
all months. Moreover, we notice that {\em Hillary Clinton}'s
attitude highly decreased in May 2016 (possibly, for example, due to
a report issued by the State Department related to Clinton's use of
private email), while October 2016 was the month with the lowest attitude value for 
Donald Trump (possibly due to the several sexual assault allegations leveled against Donald Trump during that period). 
The top-3 months of {\em high attitude} in 2016 are [February, December, April] for Donald Trump, and [February, March, April] for Hillary Clinton, while the corresponding top-3 months of {\em low attitude} are [October, June, August] for Donald Trump, and [May, October, August] for Hillary Clinton.

In general, we notice that the attitude values are relatively small and close to zero. This is due to the very big number of tweets with no sentiment (almost half of the tweets).

\subsubsection*{Entity Sentimentality}
Figure \ref{fig:attSentContr} (middle) depicts the sentimentality of {\em Donald Trump} and {\em Hillary Clinton} within 2016.
We notice that for the
majority of months the tweets mentioning {\em Donald Trump} are a
bit more sentimental than those mentioning {\em Hillary Clinton}.

October seems to be one of the most \q{sentimental} months for both {\em Donald Trump} and {\em Hillary Clinton}, possibly due to the several revelations that were uncovered for both candidates the period before the US presidential election (held on November 8).

\subsubsection*{Entity Controversiality} 
Figure \ref{fig:attSentContr} (right) shows the controversiality of {\em
Donald Trump} and {\em Hillary Clinton} within 2016 (using $\delta =
2.0$). We notice that {\em Donald Trump} induces more controversial
discussions in Twitter than {\em Hillary Clinton}, while February was
his most \q{controversial} month, probably because of his references
to some debatable topics during his campaign trail.
It is interesting also that {\em Hillary Clinton}'s controversiality
has an exponential increment from September to December 2016 (the period before, during, and after the US presidential election).

As regards their {\em top-$K$ time periods of high/low controversiality} (Formulas \ref{eq:topKcontrHigh} and \ref{eq:topKcontrLow}),
the top-3 months of {\em high controversiality} in 2016 are [February, March, December] for {\em Donald Trump} and [December, March, February] for {\em Hillary Clinton}, while the corresponding top-3 months of {\em low controversiality} are [October, August, June] for {\em Donald Trump} and [May, June, August] for {\em Hillary Clinton}.

\begin{figure*}[t]
    \centering
    \includegraphics[width=5.0in]{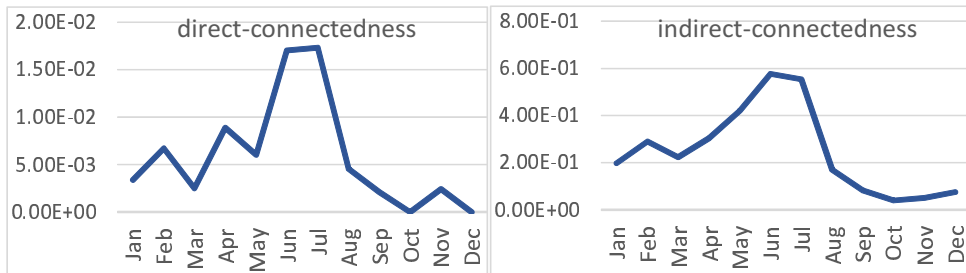}
    \caption{Direct (Formula \ref{eq:conT}) and indirect (Formula \ref{eq:conE}) connectedness of {\em \q{Alexis Tsipras}} with {\em \q{Greek withdrawal from the eurozone}} in 2015.}
    \label{fig:ent2entTsipras}
\end{figure*}

\begin{figure}[t]
    \centering
    \includegraphics[width=2.5in]{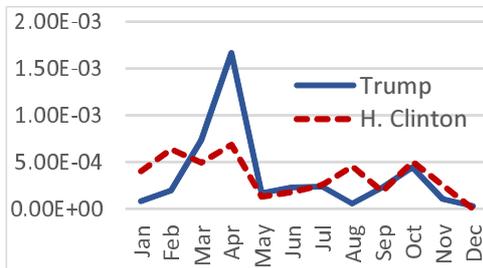}
    \caption{Direct connectedness (Formula \ref{eq:conT}) of {\em \q{Donald Trump}} and {\em \q{Hillary Clinton}} with {\em \q{Abortion}} in 2016.}
    \label{fig:ent2entAbortion}
\end{figure}

\subsubsection*{Entity-to-Entity Connectedness} 
Figure \ref{fig:ent2entTsipras} depicts the connectedness of {\em Alexis
Tsipras} with the concept {\em Greek withdrawal from the eurozone}
within 2015. We notice that these two entities are highly connected
in June and July, while after August, their connectedness is very
close to zero. Indeed, important events related to Greece's debt crisis
took place in June and July 2015, including the bank holiday, the
capital controls and the Greek bailout referendum. 

Likewise, Figure
\ref{fig:ent2entAbortion} shows the connectedness of both {\em Donald
Trump} and {\em Hillary Clinton} with the concept {\em Abortion} in
2016. Here we notice that the connectedness is almost constant for
{\em Hillary Clinton}, while for {\em Donald Trump}, there is a very
large increment in March and April. During these two months, Donald Trump made several anti-abortion comments, like that \textit{\q{there has to be some form of punishment}} for women who have abortions\footnote{\url{https://www.nytimes.com/2016/03/31/us/politics/donald-trump-abortion.html} (August 30, 2018)}.

\subsubsection*{Entity $k$-Network} 
\begin{figure*}[t]
    \centering
    \includegraphics[width=4.5in]{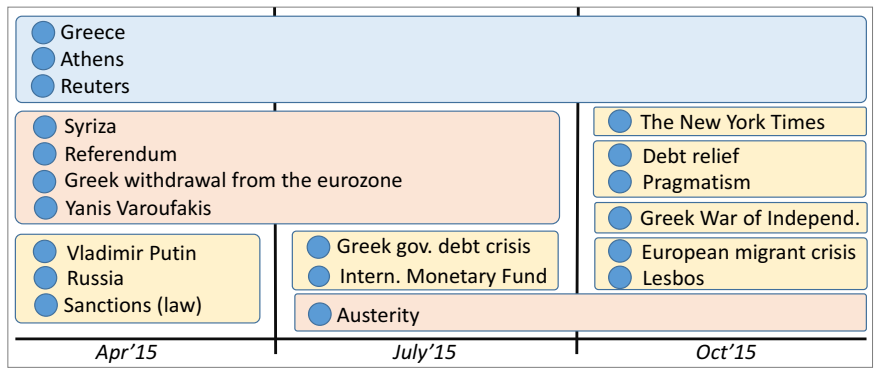}
    \caption{10-Network of {\em Alexis Tsipras} in April, July and October 2015.}
    \label{fig:Tsipras10network}
\end{figure*}

Figure \ref{fig:Tsipras10network} shows the 10-Network of {\em Alexis
Tsipras} in three different time periods (April, July and October,
2015). We notice that there are three general entities that exist in
all time periods ({\em Greece}, {\em Athens}, {\em Reuters}). For
April and July, we notice that the 10-Network contains 4 common
entities ({\em Syriza}, {\em Referendum}, {\em Greek withdrawal from
the eurozone}, and {\em Yanis Varoufakis}), while for July and
October, {\em Austerity} is the only common entity
(probably related to the approval of strict measures required by the creditors). For
April, the 10-Network contains three entities related to Russia (due
to Tsipra's visit in Moscow to meet Russian president Vladimir
Putin), while for October, it contains two entities related to
European migrant crisis (probably due to Tsipra's visit in Lesvos island).

\subsubsection*{Entity Positive \& Negative $k$-Networks} 

In Figure \ref{fig:attSentContr} (left), we saw that the attitude towards {\em Donald Trump} highly decreased in October 2016. To understand this decrement, we can inspect his {\em Negative $k$-Network} during the same period, i.e., the entities that co-occur with {\em Donald Trump} in tweets with strong average negative attitude during October 2016. Table \ref{tab:negNetworkExample} shows the results for $k=5$ and $\delta = 2.0$. 
We notice that the top-5 list contains entities related to important events that happened during this period and which are related to {\em Donald Trump}, including {\em Iraq War} (Donald Trump said that he opposed Iraq War from the start, however there appeared audio evidence of him saying he supported it), {\em Bill Clinton} (Donald Trump appeared with Bill Clinton accusers before a debate), {\em Toddler} (Donald Trump brought a toddler to the stage during a campaign rally), and {\em Central Park} (related to the Central Park jogger case: Donald Trump declared that the Central Park Five were guilty).

In Figure \ref{fig:attSentContr} (left), we also notice that the attitude towards {\em Donald Trump} highly increased in November-December 2016.
Table \ref{tab:posNetworkExample} shows the corresponding {\em Positive 5-Network} of {\em Donald Trump} for this time period (using again $\delta = 2.0$).
The top-5 list contains entities related to Donald Trump's election ({\em Cold open}, {\em God Bless America}, {\em Cheers}, and {\em Excite}) as well as the entity {\em Henry Kissinger} with whom Donald Trump met on November 17.

\begin{table}
  \caption{Negative 5-Network of Donald Trump in October 2016 (using $\delta = 2.0$).}
  \label{tab:negNetworkExample}
  \centering
  \begin{tabular}{ll}
    \toprule
    Rank&Entity\\
    \midrule
    1   & Iraq War  \\ 
	2   & Bill Clinton \\ 
	3   & Toddler \\ 
	4   & Embarrassment \\ 
	5   & Central Park \\ 
  \bottomrule
\end{tabular}

\vspace{5mm}

  \caption{Positive 5-Network of Donald Trump in November-December 2016 (using $\delta = 2.0$).}
  \label{tab:posNetworkExample}
  \centering
  \begin{tabular}{ll}
    \toprule
    Rank&Entity\\
    \midrule
    1   & Cold open  \\ 
	2   & God Bless America \\ 
	3   & Henry Kissinger \\ 
	4   & Cheers \\ 
	5   & Excite \\ 
  \bottomrule
\end{tabular}
\end{table}

\subsection{Limitations and Problems}

Although the proposed analysis approach is generic
and can be applied over different types of social media archives,
it is clear that the quality of the generated data depends
on the quality of the input data.
Twitter, for example, provides 1\% random sample, which though
is subject to bias, fake news and possibly other adversarial attacks.
In our case study,
although we remove spam, we do not take similar actions to deal with bias
and other data peculiarities.
This also means that high profile entities might occupy a big volume in the archive,
whereas long-tail entities might be underrepresented or not represented at all.

Except for the quality of the original data, the different preprocessing steps (spam removal, entity linking, sentiment analysis) are also prone to errors.
This means that the data produced by the proposed measures are also prone to errors.
For instance, regarding the entity linking task, selecting a low threshold for the confidence score of the extracted entities can result in many false annotations (disambiguation errors),
which in turn can affect the quality and reliability of the produced time-series. 
For the case of {\em Entity k-Networks} in particular, one may get some unexpected and surprising results due to disambiguation errors. 
For instance, the {\em Negative 10-Network} of Donald Trump for October 2016 returns the entity {\em Harrow (tool)}, while that of Hillary Clinton for May 2016 returns the entity {\em Clueless (film)}. Both these two entities have been incorrectly disambiguated  by FEL.


\section{Conclusion}
\label{sec:conclusion}

We have proposed an entity-centric and multi-aspect approach to analyze social media archives. For this, we defined a set of measures that allow studying how entities are reflected in social media in different time periods as well as how entity-related information evolves over time and also with respect to other entities. 
The proposed measures enable the temporal analysis of an entity in terms of its: {\em popularity} (how much discussion it generates), {\em attitude} (predominant sentiment towards the entity), {\em sentimentality} (magnitude of sentiment towards the entity), {\em controversiality} (whether there is a consensus about the sentiment towards the entity), {\em connectedness} to another entity (how strong is its connection to another entity), and {\em network} (strongly connected entities).

We believe that such a multi-aspect analysis approach is the first step towards more advanced and meaningful exploration of social media archives, while it can facilitate research in a variety of fields, such as data science, sociology, and digital humanities.

As part of our future work, we plan to exploit the rich amount of generated data for {\em prediction} of entity-related features. In particular, given an entity, our focus will be on how we can predict future values of the proposed measures (e.g., popularity or attitude in a given horizon) \cite{saleiro2016learning}. We also intend to study approaches on {\em understanding} and {\em representing} the dynamics of such evolving entity-related information, using for instance an RDF-based approach \cite{DBLP:conf/semweb/RoussakisCSFS15}.
Another interesting direction is the exploitation of the entity-relation measures on the related problems of \textit{time-aware entity relatedness} \cite{mohapatra2018timeaware} and \textit{event timeline summarisation} \cite{tran2015balancing}.

\begin{acknowledgements}
The work was partially funded by the European Commission for the ERC Advanced Grant ALEXANDRIA (No. 339233) and by the German Research Foundation (DFG) project OSCAR (Opinion
Stream Classification with Ensembles and Active leaRners).
\end{acknowledgements}

\bibliographystyle{spbasic}
\balance
\bibliography{IJDL2018__BIB}

\end{document}